\documentclass[aps,prl,preprint,superscriptaddress]{revtex4-1}

\usepackage{color, amssymb, graphicx, amsfonts,epstopdf }
\usepackage{multirow,amssymb,amsbsy,amsmath,epstopdf}
\usepackage{hyperref}
\usepackage{graphicx}
\usepackage{verbatim}
\usepackage{bm}
\usepackage{bbold}

\newcommand{\ket}[1]{\ensuremath{\left|#1\right\rangle}}

\newcommand{\abs}[1]{\ensuremath{\left| #1 \right|}}

\hypersetup{colorlinks=true, citecolor=blue}

\begin{document}


\title{Measuring a Dynamical Topological Order Parameter in Quantum Walks}


\author{Xiao-Ye Xu}
\author{Qin-Qin Wang}
\affiliation{CAS Key Laboratory of Quantum Information, University of Science and Technology of China, Hefei 230026, People's Republic of China}
\affiliation{CAS Center For Excellence in Quantum Information and Quantum Physics, University of Science and Technology of China, Hefei 230026, People's Republic of China}
\author{Markus Heyl}
\affiliation{Max-Planck-Institut f\"ur Physik komplexer Systeme, N\"othnitzer Stra{\ss}e 38, D-01187 Dresden, Germany}
\author{Jan Carl Budich}
\affiliation{Institute of Theoretical Physics, Technische Universit\"at Dresden, 01062 Dresden, Germany}
\author{Wei-Wei Pan}
\author{Zhe Chen}
\author{Munsif Jan}
\author{Kai Sun}
\author{Jin-Shi Xu}
\author{Yong-Jian Han}
\email{smhan@ustc.edu.cn}
\author{Chuan-Feng Li}
\email{cfli@ustc.edu.cn}
\author{Guang-Can Guo}
\affiliation{CAS Key Laboratory of Quantum Information, University of Science and Technology of China, Hefei 230026, People's Republic of China}
\affiliation{CAS Center For Excellence in Quantum Information and Quantum Physics, University of Science and Technology of China, Hefei 230026, People's Republic of China}


\date{\today}

\begin{abstract}
Quantum processes of inherent dynamical nature, such as quantum walks (QWs), defy a description in terms of an equilibrium statistical physics ensemble.
Up to now, it has remained a key challenge to identify general principles behind the underlying unitary quantum dynamics.
Here, we show and experimentally observe that split-step QWs admit a characterization in terms of a dynamical topological order parameter (DTOP).
This integer-quantized DTOP measures, at a given time, the winding of the geometric phase accumulated by the wave-function during the QW.
We observe distinct dynamical regimes in our experimentally realized QWs each of which can be attributed to a qualitatively different temporal behavior of the DTOP.
Upon identifying an equivalent many-body problem, we reveal an intriguing connection between the nonanalytic changes of the DTOP in QWs and the occurrence of dynamical quantum phase transitions.
\end{abstract}


\maketitle


Coherence in quantum dynamics is at the heart of fascinating phenomena beyond the realm of classical physics, such as quantum interference effects\,\cite{Anderson1958}, entanglement production\,\cite{Islam2015,Kaufman2016}, and geometric phases\,\cite{Simon1983,Berry1984,JinShi2016}. Yet, the identification of general principles behind the inherent nonequilibrium nature of unitarily evolved quantum states still accommodates central open questions\,\cite{Eisert2015}, that we experimentally address in the context of quantum walks (QWs) below\,\cite{Aharonov1993}. QWs provide a powerful and flexible platform to experimentally realize and probe coherent quantum time evolution far from thermal equilibrium. As opposed to classical random walks, QWs are characterized by quantum superpositions of amplitudes rather than classical probability distributions. This genuine quantum character has already been harnessed in various fields of physics, ranging from the design of efficient algorithms in quantum information processing\,\cite{Childs2009,Childs2013}, observation of correlated dynamics\,\cite{Schreiber2012,Sansoni2012,Poulios2014,Preiss2015} and Anderson localization\,\cite{Schreiber2011,Crespi2013}, to the realization of exotic physical phenomena in the context topological phases\,\cite{Kitagawa2010a,Kitagawa2012a,Kitagawa2012b,Obuse2011,Obuse2015,Asboth2012,Asboth2013,Asboth2014,Rakovszky2015,Cedzich2016b,Cardano2016,Cardano2017,Xiao2017,Zhan2017,Ramasesh2017,Flurin2017,Xu2018}.

In this work, we report on the direct observation of a dynamical topological order parameter (DTOP) that provides a \emph{dynamical} characterization of QWs. To this end, we realize a split-step QW in a photonic system using the framework of time multiplexing. We reconstruct the time-evolved state along the complete trajectory consisting of up to 10 time steps. Our ability to access the full set of quantum amplitudes is essential for our dynamical classification. This is because the observed DTOP is defined as a phase winding number $\omega_D(t)$ in momentum-space, namely of the so called Pancharatnam geometric phase (PGP)\,\cite{Pancharatnam1956,Samuel1988}. Quite remarkably, dynamical transitions between topologically distinct classes of QWs can be uniquely distinguished experimentally by the observed time-dependent behavior of $\omega_D(t)$: For a quench between two systems with the same topological character, we find $\omega_D(t)=0$ for all time steps; instead, for a quench between two topologically different systems, $\omega_D(t)$ also starts at $\omega_D(t=0)=0$, but monotonously changes its value at certain critical times. Generalizing these observations, we establish a unique relation between the behavior of $\omega_D(t)$ and the change over a parameter quench in the topological properties of an effective Floquet Hamiltonian that stroboscopically describes the QW.

While the QW in our experiment realizes the dynamics of a single quantum particle, we establish an underlying many-body context which explains the points at which the DTOP $\omega_D(t)$ changes non-analytically in terms of a dynamical critical phenomenon.
To this end, we map the superposition of Bloch-waves realized in the QW to a product-state of a corresponding fermionic many-body system.
Thereby, an intriguing analogy between our present experiment and the notion of dynamical quantum phase transitions (DQPTs) occurring in the unitary evolution of quenched many-body system is revealed.
With our observation of a bulk DTOP, our present work complements the recent measurement of topologically protected boundary modes in QWs\,\cite{Kitagawa2012b}, thus providing an important step towards a comprehensive understanding of the role of topology in quantum dynamics.

~\\
\noindent
{\bf\large Quantum walk setup}
~\\

\begin{figure*}
    \centering
    \includegraphics[width =0.95\textwidth]{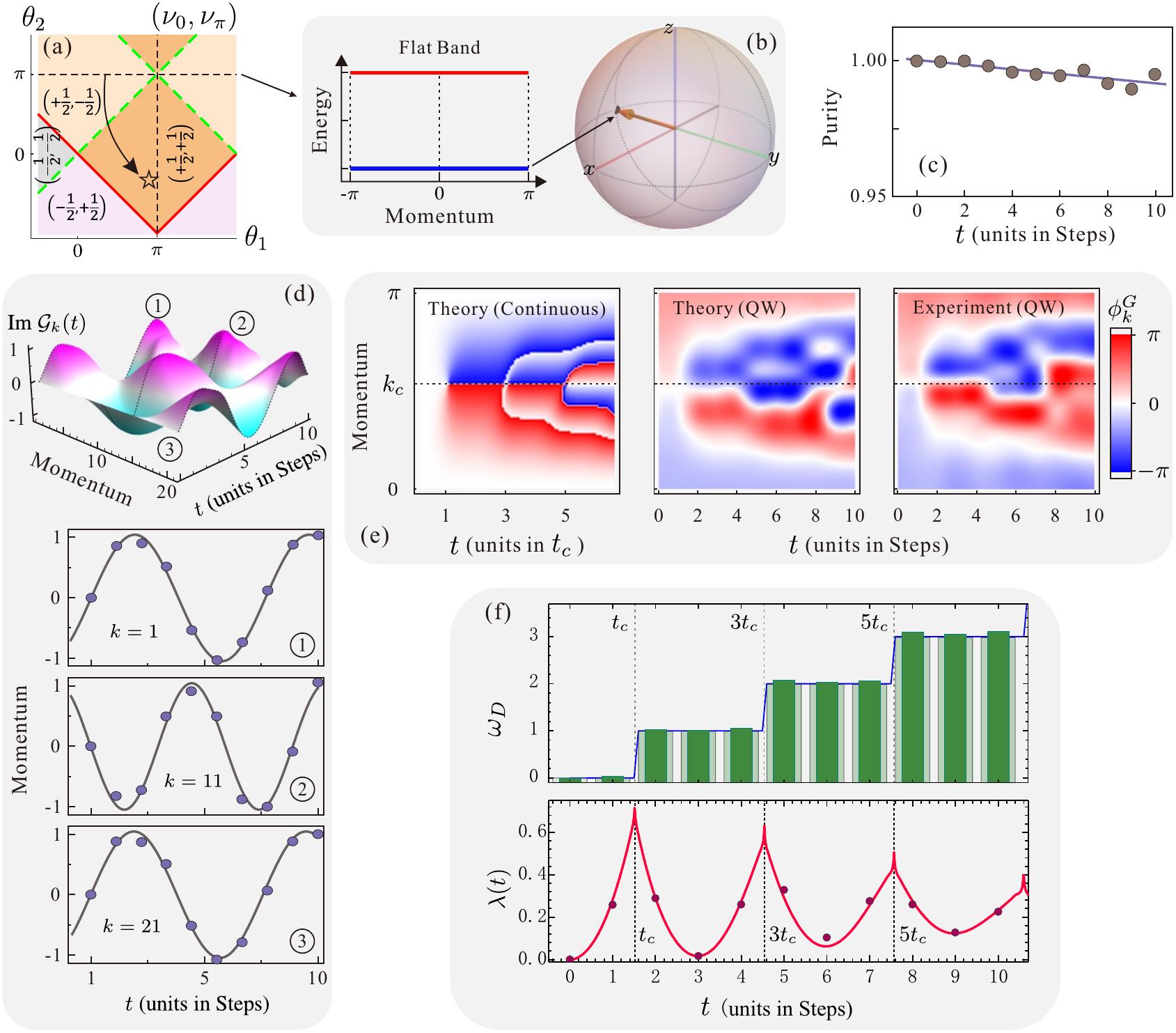}
    \caption{    }
    \label{fig:s0}
\end{figure*}

\noindent In this experiment we realize a photonic split-step QW\,\cite{Kitagawa2010a,Kitagawa2012a,Kitagawa2012b} building upon a recently introduced platform, which is based on time multiplexing and on using birefringent crystals to generate effective spin-orbit couplings~\cite{Xu2018}. For alternative implementations of QWs see Ref.\,\cite{Wang2014}. Here, we employ the two orthogonal polarizations, horizontal and vertical, as the internal coin space represented in the following as a pseudo-spin $\mu=\uparrow,\downarrow$. At each time step we repeat an identical sequence of four operations to manipulate the walker. First, a rotation $\hat R(\theta_1)$ in the internal pseudo-spin space with a tunable angle $\theta_1$ is realized via a half wave plate (HWP). This is followed by a conditional shift $\hat T_\uparrow$ of the walker to the neighboring lattice site to the right provided its internal state is $\uparrow$, achieved through a birefringent crystal. Afterwards, we perform another rotation $\hat R(\theta_2)$ with an angle $\theta_2$ and a further conditional shift $\hat T_\downarrow$, where this time the walker moves one lattice site to the left provided its internal state is $\downarrow$. Probing the dynamics stroboscopically after every completed step of the QW realizes a periodic Floquet evolution where the unitary time evolution operator $\hat U$ for one cycle is given by $\hat{U}(\theta_1,\theta_2) = \hat{T}_\downarrow\hat{R}(\theta_2)\hat{T}_\uparrow\hat{R}(\theta_1)$. Initially we prepare the photonic walker in a localized state on a given lattice site, $x=0$ say, with a tunable superposition of $\uparrow$ and $\downarrow$ in the coin space. In our experimental realization we are able to fully reconstruct the quantum state $|\Psi_t\rangle$ in the subsequent evolution of the walker (see methods)
\begin{equation}
	|\Psi_t\rangle =\sum_{x,\mu} \psi_t(x,\mu) |x \mu\rangle,
	\label{eq:defPsi_t}
\end{equation}
where $x\in\mathbb{Z}$ denotes the spatial point on the one-dimensional lattice and the quantum number $\mu=\uparrow,\downarrow$ the internal coin space.
In particular, we get access to the state amplitudes $\psi_t(x,\mu)$ at each of the up to $10$ time quench steps we study in this experiment. %
The stroboscopic evolution of our periodically time-dependent system
is determined by the associated Floquet Hamiltonian $\hat H_F(\theta_1,\theta_2)$ defined via $\hat U(\theta_1,\theta_2) = e^{-i H_F(\theta_1,\theta_2)}$.
For the split-step QW, $\hat H_F(\theta_1,\theta_2)=\int_{-\pi}^{\pi}dk H^k_F(\theta_1,\theta_2)$ is analogous to the Hamiltonian characterizing electrons in a solid with two bands, where $k$ denotes the conserved lattice momentum\,\cite{Su1979,Hasan2010}. From this perspective, this QW can exhibit interesting topological properties in the sense that the corresponding ground state represents a topological insulator, the phase diagram of $\hat H_F(\theta_1,\theta_2)$ is shown in Fig.\,\ref{fig:s0}(a). While QWs describe an inherently nonequilibrium dynamical process, signatures of these quasi-equilibrium topological properties have been observed experimentally, e.g. via the concomitant topological edge states\,\cite{Kitagawa2012b}.

\begin{figure*}
    \centering
    \includegraphics[width=0.95\textwidth]{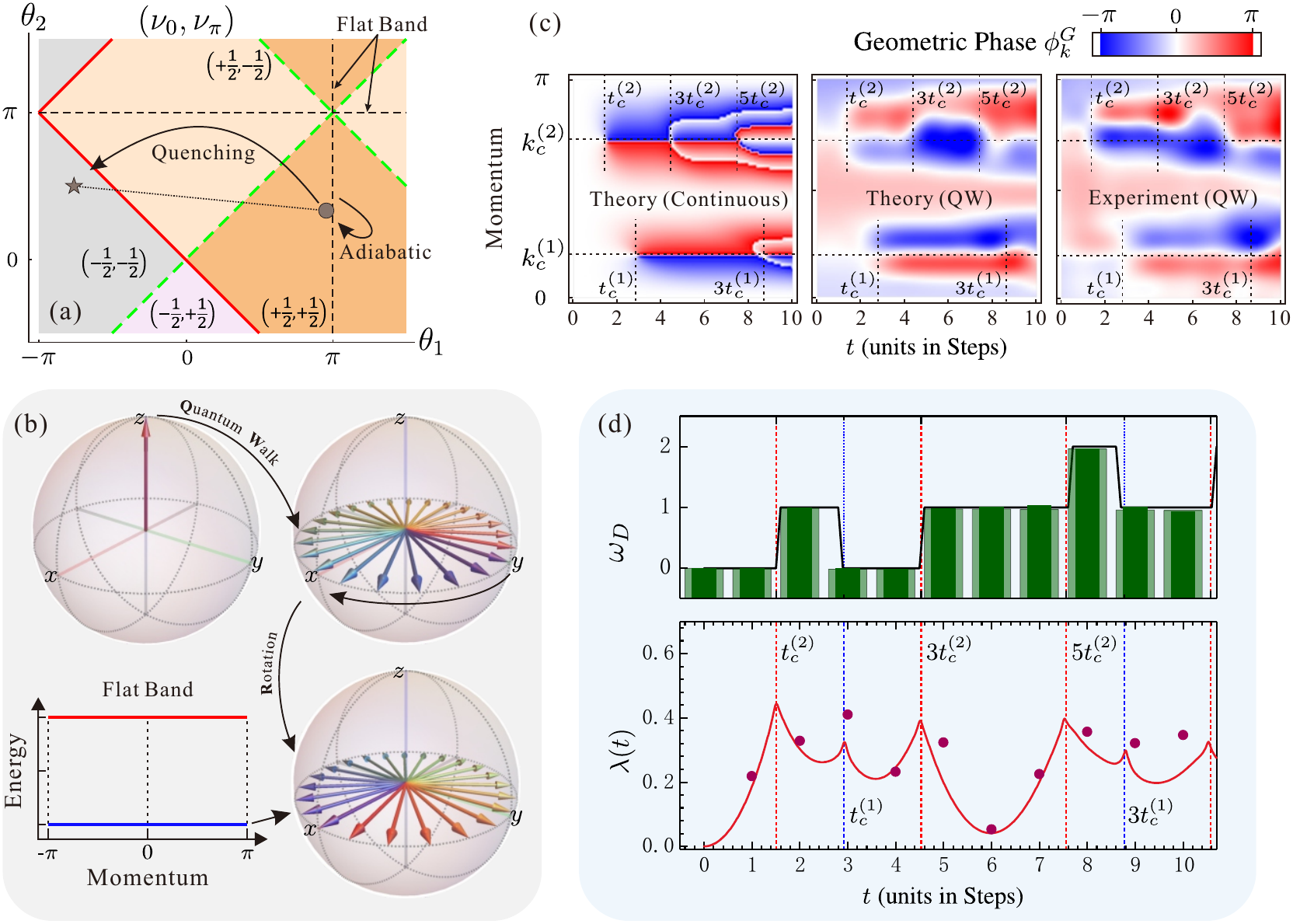}
    \caption{\textbf{Observation of the DQPT in a quench between two different non-trivial topological phases.} The strategy in phase diagram is given in (a). We start the quench in phase ($+\frac{1}{2},+\frac{1}{2}$) with $\theta_1 = 8.6\pi/9$ and $\theta_2=\pi/3$ and end it in phase ($-\frac{1}{2},-\frac{1}{2}$) with $\theta_1 = -7\pi/9$ and $\theta_2=\pi/2$. Initial state is prepared via adiabatic evolution starting from a ground state of Hamiltonian with flat band ($\theta_1 = \pi$). (b) shows the scheme for preparing the ground state of a Hamiltonian with flat band in topological non-trivial phase. The initial state is prepared via an adiabatic evolution starting from the ground state. (c) the PGP and (d) the DTOP. In this scenario, two critical times are observed. The rate function $\lambda (t)$ is shown in the bottom of (d) with the experimental results given by points.}
    \label{fig:s3}
\end{figure*}

The purpose of our present work is to go beyond such a quasi-equilibrium picture and to characterize the dynamics of the QW through a DTOP. To this end, we initially prepare the walker at $t=0$ as a wave packet localized at $x=0$ with $|\Psi_0\rangle = \sum_{\mu} \psi_0(0,\mu) |0 \mu \rangle$. We choose the superposition in the coin space such that $|\Psi_0\rangle$ represents a single-particle eigenstate in the lower of the two bands of an initial Floquet Hamiltonian $H_F^i$, that we can also implement dynamically in our setup. This is possible whenever $H_F^i$ exhibits flat bands, as it can be realized both for the case where $H_F^i$ is topologically trivial or nontrivial, see Fig.\,\ref{fig:s0}(b) and Fig.\,\ref{fig:s3}(b). Afterwards, we evolve the system according to the chosen split-step QW characterized by $H_F$, sequencing and monitoring the full non-equilibrium dynamics of the wave-function. This protocol can be interpreted as a quantum quench from $H_F^i$ to $H_F$, which, as detailed below, we can identify with a quench in a corresponding many-body system. Even though the ground state of $H_F$ cannot be reached in a QW, from the observed dynamics of the DTOP we obtain information about its topological properties.

~\\
\noindent
{\bf\large Dynamical topological order parameter}
~\\~\\
\noindent
For the definition of the DTOP it is essential that we have experimental access to the full amplitudes $\psi_t(x,\mu)$ including also the phase information.
In this sense, the dynamical characterization we propose relies crucially on the quantum nature of the QW.
The DTOP is defined through a lattice momentum dependent PGP $\phi_k^G(t)$ extending the concept of Berry's geometric phase to non-adiabatic and non-cyclic dynamics, that is naturally realized in our QW experiment.
Specifically, $\phi_k^G(t)$ measures a gauge invariant and geometric content of the acquired phase during evolution at a given lattice momentum $k$.
In formal terms, let us expand the state at a given time step $t$ not in the real-space basis $|x\mu\rangle$ as in Eq.\,(\ref{eq:defPsi_t}) but rather in the lattice momentum basis via $|\Psi_t\rangle = \int_{-\pi}^\pi dk \, | \psi_t(k)  \rangle$ with $|\psi_t(k)\rangle = \sum_{\mu} \psi_t(k,\mu) | k\mu\rangle$ and $\psi_t(k,\mu)$ the Fourier transform of $\psi_t(x,\mu)$.
The acquired phase $\phi_k(t)$ relative to the initial condition at a given $k$ and time step $t$ can be obtained from a polar decomposition of the Loschmidt amplitude $\mathcal{G}_k(t) = \langle \psi_0(k) | \psi_t(k)\rangle = r_k(t) e^{i\phi_k(t)}$.
Importantly, $\phi_k(t)$ contains a gauge invariant part $\phi_k^G(t) = \phi_k(t)-\phi_k^\mathrm{dyn}(t)$, called the Pancharatnam geometric phase (PGP), after subtracting a dynamical contribution, which in our case of a sudden quench is given by $\phi_k^\mathrm{dyn}(t)=-t \langle \psi_t(k) | H_F | \psi_t(k) \rangle $.
With this we can now define the DTOP $\omega_D(t)$ as an integer-valued winding number associated with $\phi_k^G(t)$:
\begin{equation}
\omega_D(t)=\frac{1}{2\pi}\int^{\pi}_0\frac{\partial{\phi^G_k(t)}}{\partial{k}} \in \mathbb{Z}.
\label{def:dtop}
\end{equation}
An analog of the DTOP in the context of quenched topological superconductors has been defined in Ref.\,\cite{Budich2016}.
The quantization of $\omega_D$ is imposed by particle-hole symmetry, which ensures that $\phi_{k=0}^G(t)=\phi_{k=\pi}^G(t)$ and consequently $\phi_k^G(t)$ forms a loop on the unit circle\,\cite{Budich2016}.
Based on full state reconstruction of $\psi_t(x,\mu)$, we measure in our experiment the acquired phase $\phi_k(t)$ and, importantly, also the dynamical one $\phi_k^\mathrm{dyn}(t)$, which allows us to map out the full momentum-dependent PGP $\phi_k^G(t)$, see Methods for technical details.
In Fig.\,\ref{fig:s0}(e) we show exemplarily for one realization of the split-step QW the experimentally obtained $\phi_k^G(t)$ along a trajectory of $10$ time steps, comparing also to the theoretically expected values.
For the first few time steps, the experimental data follows closely the ideal theoretical one.
At later times deviations become visible, which we trace back mainly to decoherence in the experiment leading to a reduction of the purity of the walker's state, which we experimentally estimate in Fig.~\ref{fig:s0}(c) .
A loss of purity of a few percent already leads to substantial changes in the details of $\phi_k^G(t)$, highlighting the accuracy required both in the implementation of the unitary dynamics as well as in the state reconstruction.
However, we find that the DTOP $\omega_D(t)$ as a dynamical topological quantum number is much more robust against the loss of purity, as we will show below.

~\\
\noindent
{\bf\large Dynamical phase diagram of the split-step quantum walk}
~\\

\noindent
In the following, we use the observed integer-valued quantum number $\omega_D(t)$ to dynamically characterize the realized quenched split-step QW.
The Floquet Hamiltonian before and after the quench is characterized by a doublet of topological invariants $(\nu_0,\nu_\pi)$ each of which can take the values $\pm1/2$ in our setup.
When simply calling a Floquet Hamiltonian topologically trivial or non-trivial, we refer to the coarser $\mathbb{Z}_2$ classification obtained from
the sign of the product $\nu_0\nu_\pi$, where sign$(\nu_0\nu_\pi)=-1$ hallmarks the trivial phase.
We start by considering a setup where the initial condition of the walker implements an eigenstate of an associated topologically trivial Floquet Hamiltonian $H_F^i=H_F(8\pi/9,\pi)$ with $(\nu_0,\nu_\pi)=(-1/2,1/2)$ and the subsequent time evolution is governed by a topologically nontrivial $H_F = H_F(8\pi/9,-\pi/3)$ with $(\nu_0,\nu_\pi)=(1/2,1/2)$ see Fig.\,\ref{fig:s0}(a).
Figure\,\ref{fig:s0}(e) shows our measured PGP $\phi_k^G(t)$ for this experimental sequence.
With this we can further obtain our DTOP $\omega_D(t)$, which we show in Fig.\,\ref{fig:s0}(f) following closely  the ideal theoretically expected values.
For the first two time steps the DTOP is consistent with $\omega_D(t)=0$.
After that, however, we observe a sudden jump of the DTOP to $\omega_D(t)=1$, and similarly, for later times, it jumps to $\omega_D(t)=2$.
Since $\omega_D(t)$ is a quantized integer this change of $\omega_D(t)$ can only happen in a non-analytic fashion, indicative of a behavior that is typically associated with phase transitions.
Below, we will show that such a relation to a dynamical analog of a phase transition can indeed be established.

\begin{figure*}
    \centering
    \includegraphics[width=0.95\textwidth]{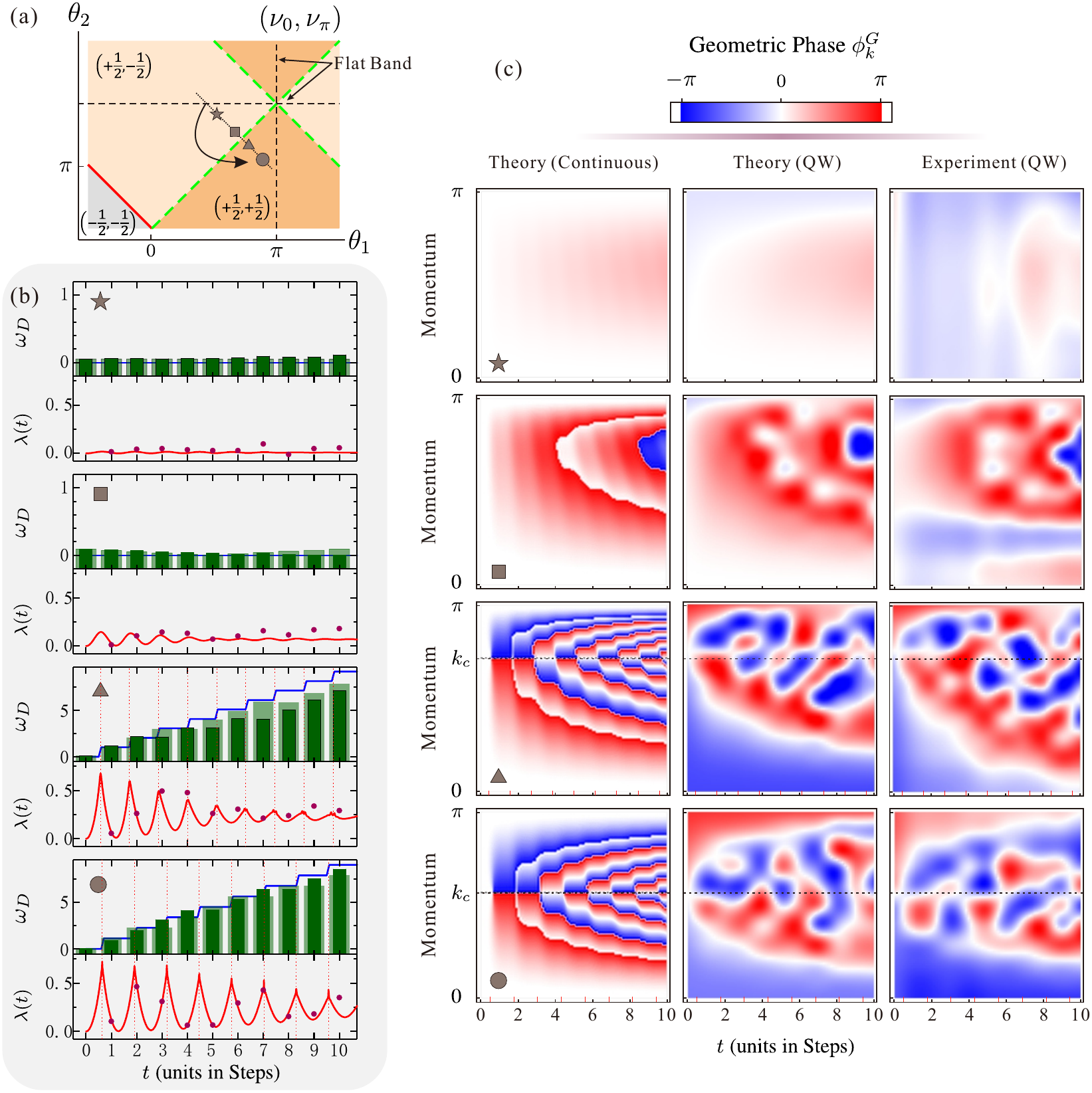}
    \caption{\textbf{Measurement of the DTOP for determining the phase boundary.} (a) shows the quenching strategies in the phase diagram: we start the quench from the flat band ($\theta_2 = \pi$) and end it in different selected Hamiltonians, two still in trivial phase and two in non-trivial phase for comparison. $\theta_1=5\pi/9~\&~\theta_2=8\pi/9$ for pentagram, $\theta_1=6\pi/9~\&~\theta_2=7\pi/9$ for square, $\theta_1=7\pi/9~\&~\theta_2=6\pi/9$ for diamond and $\theta_1=8\pi/9~\&~\theta_2=5\pi/9$ for circle, which form a line crossing the phase boundary. In (b), we show the corresponding rate function $\lambda (t)$ and the DTOP $\omega_D$ as a function of time. Lines are predicted in continuous calculation. Points and opaque bars are experimental results. Transparent bars are predictions from the simulation of QW. (c) shows the density plots of the corresponding PGP for each cases.}
    \label{fig:s1}
\end{figure*}

We now study the dynamics of the QW not only for a fixed parameter set, but rather along a line in parameter space upon keeping the initial condition fixed as specified in Fig.\,\ref{fig:s1}(a).
The time evolution of the DTOP for the different sets $(\theta_1,\theta_2)$ is shown in Fig.\,\ref{fig:s1}(b).
For $\theta_1=5\pi/9,~\theta_2=8\pi/9$, indicated by a star and which is closest in distance to the initial condition, we observe
that the DTOP $\omega_D(t)=0$ vanishes along the full trajectory.
We find the same behavior also for $\theta_1=6\pi/9,~\theta_2=7\pi/9$ (square symbol), represents a qualitatively different dynamical regime compared to the case that has been observed in Fig.\,\ref{fig:s0}(f).
However, as soon as our parameter quench crosses the boundary between the two Floquet regimes characterized by $(\nu_0,\nu_\pi)=(1/2,-1/2)$ and $(\nu_0,\nu_\pi)=(1/2,1/2)$, we recover the jumps in $\omega_D(t)$ at successive times with an overall monotonously increasing envelope when turning to the next parameter sets $\theta_1=7\pi/9,~\theta_2=6\pi/9$ (triangle
symbol) and $\theta_1=8\pi/9,~\theta_2=5\pi/9$ (circle symbol),
respectively.
According to these observations, we can identify at this point two qualitatively different dynamical phases as characterized by the temporal behavior of the DTOP.

We find, however, that there exists also a third phase characterized by yet another behavior of the $\omega_D(t)$.
For this purpose we study the DTOP for a different initial condition, for which the hypothetical ground state of the Floquet Hamiltonian $H_F^i$ would be of topological nature with $(\nu_0,\nu_\pi)=(+1/2,+1/2)$ see Fig.\,\ref{fig:s3}(a).
Upon time-evolving with an $H_F$ corresponding to a different topological phase, we observe again that the DTOP changes its value at a sequence of points in time.
Different from the previous cases, however, we observe that the DTOP can behave nonmonotonously over time.
By drawing an analogy between the realized QW and an equivalent quantum many-body problem, we will be able to explain the three observed dynamical phases in terms of a DQPT below.

\begin{figure*}
    \centering
    \includegraphics[width=0.85\textwidth]{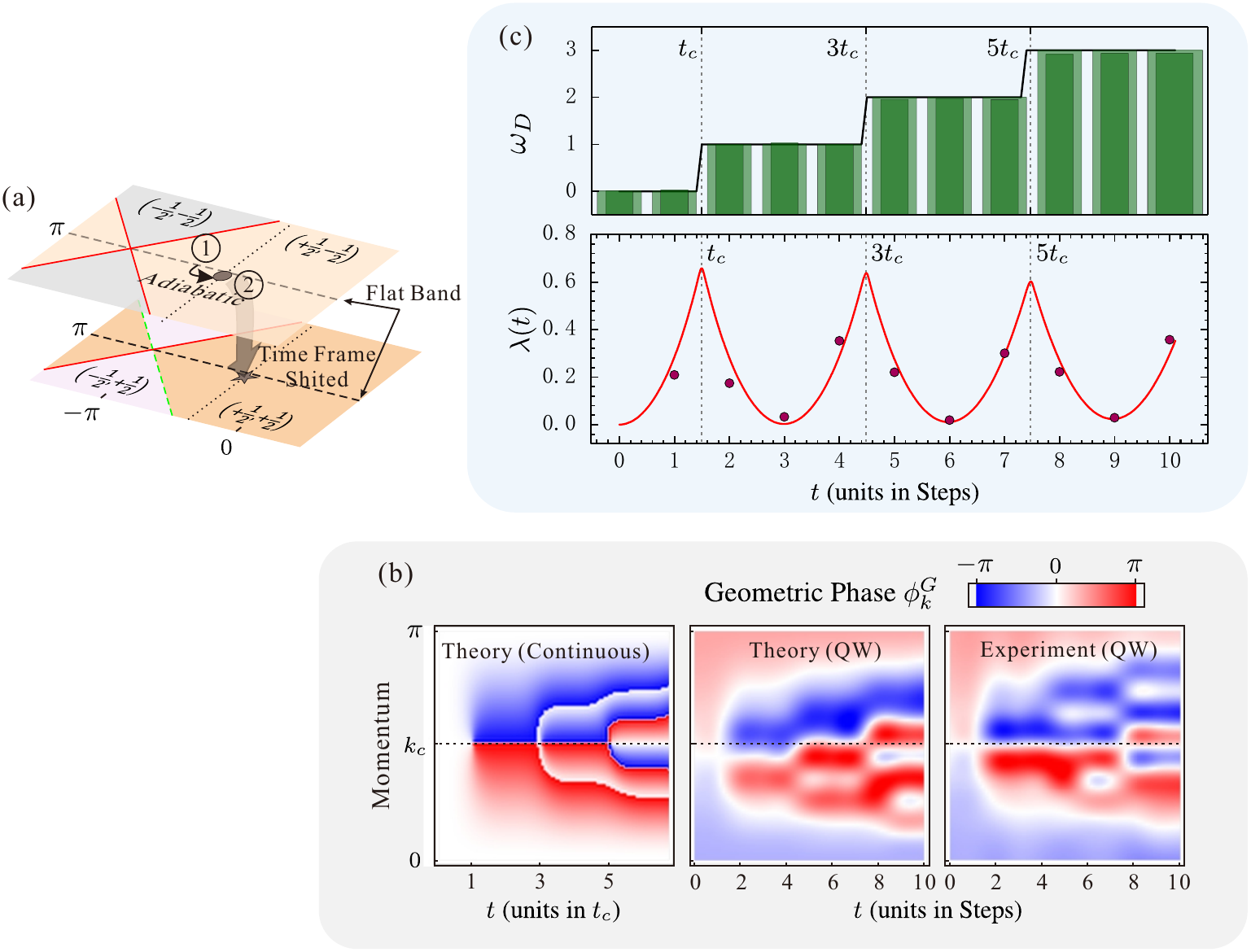}
    \caption{\textbf{Observation of the DQPT in a quench by shifting the time frame}. The strategy in phase diagram is presented in (a). We initialize the system in a ground state of a given Hamiltonian $H(\theta_1,\theta_2)$ with $\theta_1 = -\pi/3,~\theta_1 = -8.8\pi/9$, which is located in trivial phase. We adopt the adiabatic evolution starting from the flat band again to prepare the initial state. With the time frame shifted, the winding of eigenvectors of the given Hamiltonian changes its features, from zero to $\pm1$. We show the PGP in (b) and the DTOP in (c).}
    \label{fig:s2}
\end{figure*}

For the complete classification of a periodically driven system, it is important to consider different time frames\,\cite{Asboth2012,Asboth2013,Asboth2014,Obuse2015,Rakovszky2015,Cedzich2016b}. To this end, besides the conventional quench realized by sudden changes of the control parameters $\theta_1$ or $\theta_2$, we now consider a modified quench protocol, i.e., a quench induced by sudden change of the time frame (see Fig.\,\ref{fig:s2}(a) for an illustration). First, we fix the time frame $U$. By performing an adiabatic evolution starting from the origin with the spinor state $\ket{\downarrow_{y}}$ (which is the superposition state of the lower band states for QWs with the parameters constrained on the dashed line in the trivial phase as shown in Fig.\,\ref{fig:s2}(a)), the system can be further initialized in the superposition state of the lower band states of a more general QW with Hamiltonian $H_\text{eff}(-\pi/3,8.8\pi/9)$ (Note this QW is still in trivial phase). Then, we suddenly change the time frame from $U$ to a non-equivalent one $U'$\,\cite{Xu2018} meanwhile keeping the parameters unchanged. Nevertheless, the effective Hamiltonians $H_F$ become different with different topological invariants upon changing time frames. The experimental results for this scenario are shown in Fig.\,\ref{fig:s2}(b) and (c). Again we observe a characteristic behavior of the DTOP, monotonously increasing in time, which corresponds to the dynamical phase shown in Fig.\,\ref{fig:s0}(f) and Fig.\,\ref{fig:s1}(b) as expected.

~\\
\noindent
{\bf\large Dynamical quantum phase transitions}
~\\

\noindent
The real-time nonanalytic behavior of the DTOP affords an intriguing analogy to the phenomenon of DQPTs\,\cite{Heyl2013,Markus2018}, which allows us to explain our observations in light of an equivalent many-body problem.
To this end, we map our QW system, for which the state is given by a coherent superposition $|\Psi_t\rangle = \int_{-\pi}^\pi dk \, | \psi_t(k)  \rangle$ of lattice-momentum modes (Bloch states), to a fermionic many-body system, whose state is given by a Slater determinant of the $| \psi_t(k)  \rangle$.
We note that this mapping requires access to all $|\psi_t(k)\rangle$, which we can approach in our setup in parallel due to the large degree of coherence.
Within the theory of DQPTs, the central object is the Loschmidt amplitude $\mathcal{G}(t)$ which for the corresponding many-body system factorizes as $\mathcal G(t)=\prod_k \mathcal{G}_k(t)$.
DQPTs are hallmarked by non-analytic points in time of the associated rate function $g(t)=-N^{-1} \log[\mathcal G(t)]$, which plays the role of a formal analog to a free energy density.
Here, $N$ denotes the number of degrees of freedom, i.e., the number of involved lattice-momentum modes.
Such DQPTs and signatures thereof have been recently observed in various systems\,\cite{Jurcevic2017,Zhang2017,Bernien2017,Flaschner2018}.

In all the figures, we have included a theoretical calculation of
$\lambda(t)=2\text{Re}[g(t)]$ for the many-body system equivalent to the respective implemented QW.
For example, in Fig.\,\ref{fig:s0}(f) the situation corresponds to a quantum quench in a two-band fermionic system from an initial topologically trivial insulating state, the ground state of $H_F^i$, to a final Hamiltonian $H_F$ exhibiting topologically nontrivial properties.

Using the analogy to the equivalent many-body system, we can further relate the equilibrium properties of the Floquet Hamiltonian $H_F$ to the dynamics of the DTOP we observe for the QW. First of all, it is shown that a jump in the DTOP always comes along with a DQPT in the considered systems\,\cite{Heyl2013,Markus2018}. This is indeed what we find in our experiment. Those times where the observed DTOP changes its topologically quantized value coincide with the critical times at which the corresponding many-body system undergoes a DQPT, as
hallmarked by a logarithmic singularity in $g(t)$.

All potential DQPTs, that can occur in the considered models, can be grouped into two classes, which give the overall classification in terms of three dynamical phases, the third one being dynamics without the occurrence of a DQPT yielding $\omega_D(t) = 0$. First, DQPTs have to occur whenever the initial and final Hamiltonians, here $H_F^i$ and $H_F$, are topologically inequivalent in the $\mathbb{Z}_2$ classification corresponding to positive or negative sign of $\nu_0\nu_\pi$\,\cite{Vajna2015}, where $\nu_0\nu_\pi>0$ refers to the topological and $\nu_0\nu_\pi<0$ to the trivial phase, respectively. In this sense, these DQPTs are topologically protected, whose data is shown in Fig.\,\ref{fig:s0}. Second, DQPTs can be accidental, without changing the product $\nu_0\nu_\pi$ thus leaving the $\mathbb{Z}_2$ classification of the static system unchanged. Notably, in our present Floquet context, such accidental DQPTs occur precisely when both $\nu_0$ and $\nu_\pi$ switch sign while leaving their product unchanged. This gives a clear topological meaning also to this second kind of DQPTs in our split-step QW setup. Remarkably, the DTOP observed in this work is capable of qualitatively distinguishing these different kinds of DQPT scenarios (cf. Fig.\,\ref{fig:s0} and Fig.\,\ref{fig:s3}).

~\\
\noindent
{\bf\large Conclusion and outlook}

\noindent
In this experiment we have provided a dynamical characterization
of split-step QWs using a DTOP - an integer-valued quantum number which measures the winding of the geometric phase in lattice momentum space. Central for our measurement of the DTOP has been the possibility to reconstruct the full wave-function of the QW state with access to the full set of quantum amplitudes including their phase information. Due to a mapping onto a quantum quench in an equivalent quantum many-body problem, we have shown that this dynamical characterization is intimately related to the phenomenon of DQPTs in unitary real-time evolution. In this way we provide a nonequilibrium perspective onto QWs, which can be understood
as a starting point towards approaching time-dependent processes from an inherently dynamical angle that goes beyond the notion of equilibrium statistical physics. With this and the mapping onto quenches in an equivalent quantum many-body system, our experiment offers a platform to study nonequilibrium dynamics in the future.
Note added: During the preparation of the manuscript, four other experimental works have been made public which address measurements of PGP or dynamical quantum phase transitions\,\cite{Guo2018,Wang2018,Smale2018,Tian2018}. Here, based on our directly obtained DOTP, we give a dynamical classification of the quenched QW, establish the relation between the temporal behavior of DTOP and the underlying quasi-equilibrium picture, and further reveal a intriguing connection between the non-analytic of the DTOP and DQPT.

~\\
\noindent
{\bf\large Methods}

\noindent
\emph{\textbf{Initial state preparation.}} Before starting the quantum walks, we prepare the system initially in a single-particle eigenstate of an effective Floquet Hamiltonian $H^i_F$, which we can finally associate to a quantum quench
in an equivalent quantum many-body problem. We proceed by distinguishing three different cases for $H^i_F$: (a) Trivial flat band Hamiltonian, (b) Topologically non-trivial flat-band Hamiltonian, (c) General Hamiltonian without flat bands. As for (a) the ground state of the flat band can be localized on a single site at the origin in real-space with the spin pointing to the any $y$-direction, e.g., $\ket{\Psi_0} = \ket{x=0\downarrow_y}$ for $H_F^i(\theta_1^i,\pi)$. The situation in scenario (b) is a bit more complicated. We first initialize the system in a state $\ket{x=0\uparrow}$. Then, we perform a full QW step with parameters $(\pi,\pi/2)$ and finally, an additional spin rotation along $\sigma_z$ axis with an angle $\pi/2$ is performed (see Fig.\,\ref{fig:s3}(b)). In this way, the system is prepared in the state $(\ket{-1\uparrow}-i\ket{0\downarrow})/\sqrt{2}$, which, in its momentum space
representation, corresponds to a superposition including all those single-particle states in the lower band of the non-trivial flat-band Hamiltonian $H^i_F(\pi,\theta_2^i)$. Case (c) is important for realizing in the equivalent many-body problem effectively quantum quenches between two inequivalent non-trivial Hamiltonians $H^i_F$ nd $H_F$ and for a quantum quench driven by the change of time frame. To achieve an initial state corresponding to a non flat-band
Hamiltonian, we start from a flat-band ground state according to (a) or (b). Then, we perform an additional step to adiabatically transfer the system into the ground
state of a general target Hamiltonian in the same phase, which is always possible due to the finite energy gap.

\noindent
\emph{\textbf{Full state reconstruction.}} Our new platform for implementing QWs allows us to access to the full wave-function including the phase information (see Ref.\,\cite{Xu2018} for a detailed discussion). In brief terms, suppose that the
system after $t$ steps of the QW is in a state $\ket{\Psi_t}$ (see Eq.\,\ref{eq:defPsi_t}). We then carry out three steps to get the complex coefficients $\psi_t(x,\mu)$: First, we perform a local projection measurement on the spin for each site and get a count set $\mathcal{S}$. Then, after shifting all of the spin-up components a step backward (by inserting an additional birefringent crystal), we perform a local projection measurement on the spin again and get another count set $\tilde{\mathcal{S}}$. Finally, based on a simulated annealing algorithm, we carry out a numerical global program to find an optimal state of the form given in Eq.\,\ref{eq:defPsi_t} which reproduces the two count sets $\mathcal{S},\tilde{\mathcal{S}}$ with the largest probability. As the number of projection bases equals $4(2N-1)$ with $N$ the lattice size, which is much greater than the number of independent variables $2(2N-1)$ in the wave-function Eq.\,\ref{eq:defPsi_t}, we can systematically improve the rank of the target state and monitor the decoherence in the experiment. With the full knowledge of $\ket{\Psi_t}$, i.e. both the amplitudes and phases of the coefficients $\psi_t(x,\mu)$, we can readily obtain the wave-function in momentum representation by performing a Fourier transform. Concretely, we perform a discrete Fourier transform separately to the spin-up and the spin-down component, and then renormalize the components for each quasi-momentum.

\noindent
\emph{\textbf{Measuring the Pancharatnam geometric phase.}}
We now provide details on how the PGP, which is at the heart of our present study, can be directly extracted from our experimental data. We focus on the PGP $\phi_k^G$ associated with a fixed lattice momentum $k$, defined via $\mathcal{G}_k(t) = \langle\psi_0(k)|\psi_t(k)\rangle = r_k(t)e^{i\phi_k(t)}$ with $\phi_k(t) = \phi_k^G(t) + \phi_k^\text{dyn}(t)$. Our direct observation of $\phi_k^G$ then results from the independent observation of the total phase $\phi_k(t)$ and the dynamical phase $\phi_k^\text{dyn}(t)$ of the time-evolved wave-function $\ket{\psi_t(k)}$ relative to the initial condition $\ket{\psi_k(0)}$. The total phase is an immediate result of our full state tomography of the time-evolved wave-function. To isolate the dynamical phase $\phi_k^\text{dyn}(t)$, we expand the initial state $\ket{\psi_t(0)} = g_k\ket{u_k^-} + e_k\ket{u_k^+}$ in the eigenbasis $\ket{u_k^\pm}$ of the final Hamiltonian $H_F$ with $\pm\epsilon_k^f$ denoting the corresponding eigenenergies. In this representation, the Loschmidt amplitude takes the form $\mathcal{G}_k(t) = (\abs{g_k}^2 + \abs{e_k}^2)\cos{(\epsilon_k^ft)} + i(\abs{g_k}^2 - \abs{e_k}^2)\sin{(\epsilon_k^ft)}$. By observing the amplitude and phase of the oscillations of $\mathcal{G}_k(t)$, we hence obtain $(\abs{g_k}^2 - \abs{e_k}^2)$ and $\epsilon_k^f$, respectively, see also Fig.\,\ref{fig:s0}(d). This determines the dynamical phase $\phi_k^\text{dyn}(t) = \epsilon_k^ft(\abs{g_k}^2 - \abs{e_k}^2)$ and thus the PGP $\phi_k^G(t)=\phi_k(t)-\phi_k^\text{dyn}(t)$.

\bibliographystyle{naturemag.bst}
\bibliography{references}

\providecommand{\noopsort}[1]{}\providecommand{\singleletter}[1]{#1}%
\begin{thebibliography}{10}
\expandafter\ifx\csname url\endcsname\relax
  \def\url#1{\texttt{#1}}\fi
\expandafter\ifx\csname urlprefix\endcsname\relax\def\urlprefix{URL }\fi
\providecommand{\bibinfo}[2]{#2}
\providecommand{\eprint}[2][]{\url{#2}}

\bibitem{Anderson1958}
\bibinfo{author}{Anderson, P.~W.}
\newblock \bibinfo{title}{Absence of diffusion in certain random lattices}.
\newblock \emph{\bibinfo{journal}{Phys. Rev.}} \textbf{\bibinfo{volume}{109}},
  \bibinfo{pages}{1492} (\bibinfo{year}{1958}).

\bibitem{Islam2015}
\bibinfo{author}{Islam, R.} \emph{et~al.}
\newblock \bibinfo{title}{Measuring entanglement entropy in a quantum many-body
  system}.
\newblock \emph{\bibinfo{journal}{Nature}} \textbf{\bibinfo{volume}{528}},
  \bibinfo{pages}{77} (\bibinfo{year}{2015}).

\bibitem{Kaufman2016}
\bibinfo{author}{Kaufman, A.~M.} \emph{et~al.}
\newblock \bibinfo{title}{Quantum thermalization through entanglement in an
  isolated many-body system}.
\newblock \emph{\bibinfo{journal}{Science}} \textbf{\bibinfo{volume}{353}},
  \bibinfo{pages}{794--800} (\bibinfo{year}{2016}).

\bibitem{Simon1983}
\bibinfo{author}{Simon, B.}
\newblock \bibinfo{title}{Holonomy, the quantum adiabatic theorem, and berry's
  phase}.
\newblock \emph{\bibinfo{journal}{Phys. Rev. Lett.}}
  \textbf{\bibinfo{volume}{51}}, \bibinfo{pages}{2167--2170}
  (\bibinfo{year}{1983}).

\bibitem{Berry1984}
\bibinfo{author}{Berry, M.~V.}
\newblock \bibinfo{title}{Quantal phase factors accompanying adiabatic
  changes}.
\newblock \emph{\bibinfo{journal}{Proc. R. Soc. A}}
  \textbf{\bibinfo{volume}{392}}, \bibinfo{pages}{45} (\bibinfo{year}{1984}).

\bibitem{JinShi2016}
\bibinfo{author}{Xu, J.-S.} \emph{et~al.}
\newblock \bibinfo{title}{Simulating the exchange of majorana zero modes with a
  photonic system}.
\newblock \emph{\bibinfo{journal}{Nat. Commun.}} \textbf{\bibinfo{volume}{7}},
  \bibinfo{pages}{13194} (\bibinfo{year}{2016}).

\bibitem{Eisert2015}
\bibinfo{author}{Eisert, J.}, \bibinfo{author}{Friesdorf, M.} \&
  \bibinfo{author}{Gogolin, C.}
\newblock \bibinfo{title}{Quantum many-body systems out of equilibrium}.
\newblock \emph{\bibinfo{journal}{Nat. Phys.}} \textbf{\bibinfo{volume}{11}},
  \bibinfo{pages}{124} (\bibinfo{year}{2015}).

\bibitem{Aharonov1993}
\bibinfo{author}{Aharonov, Y.}, \bibinfo{author}{Davidovich, L.} \&
  \bibinfo{author}{Zagury, N.}
\newblock \bibinfo{title}{Quantum random walks}.
\newblock \emph{\bibinfo{journal}{Phys. Rev. A}} \textbf{\bibinfo{volume}{48}},
  \bibinfo{pages}{1687--1690} (\bibinfo{year}{1993}).

\bibitem{Childs2009}
\bibinfo{author}{Childs, A.~M.}
\newblock \bibinfo{title}{Universal computation by quantum walk}.
\newblock \emph{\bibinfo{journal}{Phys. Rev. Lett.}}
  \textbf{\bibinfo{volume}{102}}, \bibinfo{pages}{180501}
  (\bibinfo{year}{2009}).

\bibitem{Childs2013}
\bibinfo{author}{Childs, A.~M.}, \bibinfo{author}{Gosset, D.} \&
  \bibinfo{author}{Webb, Z.}
\newblock \bibinfo{title}{Universal computation by multiparticle quantum walk}.
\newblock \emph{\bibinfo{journal}{Science}} \textbf{\bibinfo{volume}{339}},
  \bibinfo{pages}{791--794} (\bibinfo{year}{2013}).

\bibitem{Schreiber2012}
\bibinfo{author}{Schreiber, A.} \emph{et~al.}
\newblock \bibinfo{title}{A 2d quantum walk simulation of two-particle
  dynamics}.
\newblock \emph{\bibinfo{journal}{Science}} \textbf{\bibinfo{volume}{336}},
  \bibinfo{pages}{55--58} (\bibinfo{year}{2012}).

\bibitem{Sansoni2012}
\bibinfo{author}{Sansoni, L.} \emph{et~al.}
\newblock \bibinfo{title}{Two-particle bosonic-fermionic quantum walk via
  integrated photonics}.
\newblock \emph{\bibinfo{journal}{Phys. Rev. Lett.}}
  \textbf{\bibinfo{volume}{108}}, \bibinfo{pages}{010502}
  (\bibinfo{year}{2012}).

\bibitem{Poulios2014}
\bibinfo{author}{Poulios, K.} \emph{et~al.}
\newblock \bibinfo{title}{Quantum walks of correlated photon pairs in
  two-dimensional waveguide arrays}.
\newblock \emph{\bibinfo{journal}{Phys. Rev. Lett.}}
  \textbf{\bibinfo{volume}{112}}, \bibinfo{pages}{143604}
  (\bibinfo{year}{2014}).

\bibitem{Preiss2015}
\bibinfo{author}{Preiss, P.~M.} \emph{et~al.}
\newblock \bibinfo{title}{Strongly correlated quantum walks in optical
  lattices}.
\newblock \emph{\bibinfo{journal}{Science}} \textbf{\bibinfo{volume}{347}},
  \bibinfo{pages}{1229--1233} (\bibinfo{year}{2015}).

\bibitem{Schreiber2011}
\bibinfo{author}{Schreiber, A.} \emph{et~al.}
\newblock \bibinfo{title}{Decoherence and disorder in quantumwalks: From
  ballistic spread to localization}.
\newblock \emph{\bibinfo{journal}{Phys. Rev. Lett.}}
  \textbf{\bibinfo{volume}{106}}, \bibinfo{pages}{180403}
  (\bibinfo{year}{2011}).

\bibitem{Crespi2013}
\bibinfo{author}{Crespi, A.} \emph{et~al.}
\newblock \bibinfo{title}{Anderson localization of entangled photons in an
  integrated quantum walk}.
\newblock \emph{\bibinfo{journal}{Nat. Photon.}} \textbf{\bibinfo{volume}{7}},
  \bibinfo{pages}{322--328} (\bibinfo{year}{2013}).

\bibitem{Kitagawa2010a}
\bibinfo{author}{Kitagawa, T.}, \bibinfo{author}{Rudner, M.~S.},
  \bibinfo{author}{Berg, E.} \& \bibinfo{author}{Demler, E.}
\newblock \bibinfo{title}{Exploring topological phases with quantum walks}.
\newblock \emph{\bibinfo{journal}{Phys. Rev. A}} \textbf{\bibinfo{volume}{82}}
  (\bibinfo{year}{2010}).

\bibitem{Kitagawa2012a}
\bibinfo{author}{Kitagawa, T.}
\newblock \bibinfo{title}{Topological phenomena in quantum walks: elementary
  introduction to the physics of topological phases}.
\newblock \emph{\bibinfo{journal}{Quantum Inf. Process.}}
  \textbf{\bibinfo{volume}{11}}, \bibinfo{pages}{1107--1148}
  (\bibinfo{year}{2012}).

\bibitem{Kitagawa2012b}
\bibinfo{author}{Kitagawa, T.} \emph{et~al.}
\newblock \bibinfo{title}{Observation of topologically protected bound states
  in photonic quantum walks}.
\newblock \emph{\bibinfo{journal}{Nat. Commun.}} \textbf{\bibinfo{volume}{3}},
  \bibinfo{pages}{882} (\bibinfo{year}{2012}).

\bibitem{Obuse2011}
\bibinfo{author}{Obuse, H.} \& \bibinfo{author}{Kawakami, N.}
\newblock \bibinfo{title}{Topological phases and delocalization of quantum
  walks in random environments}.
\newblock \emph{\bibinfo{journal}{Phys. Rev. B}} \textbf{\bibinfo{volume}{84}},
  \bibinfo{pages}{195139} (\bibinfo{year}{2011}).

\bibitem{Obuse2015}
\bibinfo{author}{Obuse, H.}, \bibinfo{author}{Asb\'{o}th, J.~K.},
  \bibinfo{author}{Nishimura, Y.} \& \bibinfo{author}{Kawakami, N.}
\newblock \bibinfo{title}{Unveiling hidden topological phases of a
  one-dimensional hadamard quantum walk}.
\newblock \emph{\bibinfo{journal}{Phys. Rev. B}} \textbf{\bibinfo{volume}{92}},
  \bibinfo{pages}{045424} (\bibinfo{year}{2015}).

\bibitem{Asboth2012}
\bibinfo{author}{Asb\'{o}th, J.~K.}
\newblock \bibinfo{title}{Symmetries, topological phases, and bound states in
  the one-dimensional quantum walk}.
\newblock \emph{\bibinfo{journal}{Phys. Rev. B}} \textbf{\bibinfo{volume}{86}},
  \bibinfo{pages}{195414} (\bibinfo{year}{2012}).

\bibitem{Asboth2013}
\bibinfo{author}{Asb\'{o}th, J.~K.} \& \bibinfo{author}{Obuse, H.}
\newblock \bibinfo{title}{Bulk-boundary correspondence for chiral symmetric
  quantum walks}.
\newblock \emph{\bibinfo{journal}{Phys. Rev. B}} \textbf{\bibinfo{volume}{88}},
  \bibinfo{pages}{121406} (\bibinfo{year}{2013}).

\bibitem{Asboth2014}
\bibinfo{author}{Asb\'{o}th, J.~K.}, \bibinfo{author}{Tarasinski, B.} \&
  \bibinfo{author}{Delplace, P.}
\newblock \bibinfo{title}{Chiral symmetry and bulk-boundary correspondence in
  periodically driven one-dimensional systems}.
\newblock \emph{\bibinfo{journal}{Phys. Rev. B}} \textbf{\bibinfo{volume}{90}},
  \bibinfo{pages}{125143} (\bibinfo{year}{2014}).

\bibitem{Rakovszky2015}
\bibinfo{author}{Rakovszky, T.} \& \bibinfo{author}{Asboth, J.~K.}
\newblock \bibinfo{title}{Localization, delocalization, and topological phase
  transitions in the one-dimensional split-step quantum walk}.
\newblock \emph{\bibinfo{journal}{Phys. Rev. A}} \textbf{\bibinfo{volume}{92}},
  \bibinfo{pages}{052311} (\bibinfo{year}{2015}).

\bibitem{Cedzich2016b}
\bibinfo{author}{Cedzich, C.} \emph{et~al.}
\newblock \bibinfo{title}{Bulk-edge correspondence of one-dimensional quantum
  walks}.
\newblock \emph{\bibinfo{journal}{J. Phys. A - Math. Theor.}}
  \textbf{\bibinfo{volume}{49}}, \bibinfo{pages}{21LT01}
  (\bibinfo{year}{2016}).

\bibitem{Cardano2016}
\bibinfo{author}{Cardano, F.} \emph{et~al.}
\newblock \bibinfo{title}{Statistical moments of quantum-walk dynamics reveal
  topological quantum transitions}.
\newblock \emph{\bibinfo{journal}{Nat. Commun.}} \textbf{\bibinfo{volume}{7}},
  \bibinfo{pages}{11439} (\bibinfo{year}{2016}).

\bibitem{Cardano2017}
\bibinfo{author}{Cardano, F.} \emph{et~al.}
\newblock \bibinfo{title}{Detection of zak phases and topological invariants in
  a chiral quantum walk of twisted photons}.
\newblock \emph{\bibinfo{journal}{Nat. Commun.}} \textbf{\bibinfo{volume}{8}},
  \bibinfo{pages}{15516} (\bibinfo{year}{2017}).

\bibitem{Xiao2017}
\bibinfo{author}{Xiao, L.} \emph{et~al.}
\newblock \bibinfo{title}{Observation of topological edge states in
  parity–time-symmetric quantum walks}.
\newblock \emph{\bibinfo{journal}{Nat. Phys.}}  (\bibinfo{year}{2017}).

\bibitem{Zhan2017}
\bibinfo{author}{Zhan, X.} \emph{et~al.}
\newblock \bibinfo{title}{Detecting topological invariants in nonunitary
  discrete-time quantum walks}.
\newblock \emph{\bibinfo{journal}{Phys. Rev. Lett.}}
  \textbf{\bibinfo{volume}{119}}, \bibinfo{pages}{130501}
  (\bibinfo{year}{2017}).

\bibitem{Ramasesh2017}
\bibinfo{author}{Ramasesh, V.~V.}, \bibinfo{author}{Flurin, E.},
  \bibinfo{author}{Rudner, M.}, \bibinfo{author}{Siddiqi, I.} \&
  \bibinfo{author}{Yao, N.~Y.}
\newblock \bibinfo{title}{Direct probe of topological invariants using bloch
  oscillating quantum walks}.
\newblock \emph{\bibinfo{journal}{Phys. Rev. Lett.}}
  \textbf{\bibinfo{volume}{118}}, \bibinfo{pages}{130501}
  (\bibinfo{year}{2017}).

\bibitem{Flurin2017}
\bibinfo{author}{Flurin, E.} \emph{et~al.}
\newblock \bibinfo{title}{Observing topological invariants using quantum walks
  in superconducting circuits}.
\newblock \emph{\bibinfo{journal}{Phys. Rev. X}} \textbf{\bibinfo{volume}{7}},
  \bibinfo{pages}{031023} (\bibinfo{year}{2017}).

\bibitem{Xu2018}
\bibinfo{author}{Xu, X.-Y.} \emph{et~al.}
\newblock \bibinfo{title}{Measuring the winding number in a large-scale chiral
  quantum walk}.
\newblock \emph{\bibinfo{journal}{Phys. Rev. Lett.}}
  \textbf{\bibinfo{volume}{120}}, \bibinfo{pages}{260501}
  (\bibinfo{year}{2018}).

\bibitem{Pancharatnam1956}
\bibinfo{author}{Pancharatnam, S.}
\newblock \bibinfo{title}{Generalized theory of interference, and its
  applications}.
\newblock \emph{\bibinfo{journal}{Proceedings of the Indian Academy of Sciences
  - Section A}} \textbf{\bibinfo{volume}{44}}, \bibinfo{pages}{247--262}
  (\bibinfo{year}{1956}).

\bibitem{Samuel1988}
\bibinfo{author}{Samuel, J.} \& \bibinfo{author}{Bhandari, R.}
\newblock \bibinfo{title}{General setting for berry's phase}.
\newblock \emph{\bibinfo{journal}{Phys. Rev. Lett.}}
  \textbf{\bibinfo{volume}{60}}, \bibinfo{pages}{2339--2342}
  (\bibinfo{year}{1988}).

\bibitem{Wang2014}
\bibinfo{author}{Wang, J.} \& \bibinfo{author}{Manouchehri, K.}
\newblock \emph{\bibinfo{title}{Physical Implementation of Quantum Walks}}.
\newblock Quantum Science and Technology (\bibinfo{publisher}{Springer-Verlag
  New York}, \bibinfo{year}{2014}).

\bibitem{Su1979}
\bibinfo{author}{Su, W.~P.}, \bibinfo{author}{Schrieffer, J.~R.} \&
  \bibinfo{author}{Heeger, A.~J.}
\newblock \bibinfo{title}{Solitons in polyacetylene}.
\newblock \emph{\bibinfo{journal}{Phys. Rev. Lett.}}
  \textbf{\bibinfo{volume}{42}}, \bibinfo{pages}{1698} (\bibinfo{year}{1979}).

\bibitem{Hasan2010}
\bibinfo{author}{Hasan, M.~Z.} \& \bibinfo{author}{Kane, C.~L.}
\newblock \bibinfo{title}{Colloquium: Topological insulators}.
\newblock \emph{\bibinfo{journal}{Rev. Mod. Phys.}}
  \textbf{\bibinfo{volume}{82}}, \bibinfo{pages}{3045--3067}
  (\bibinfo{year}{2010}).

\bibitem{Budich2016}
\bibinfo{author}{Budich, J.~C.} \& \bibinfo{author}{Heyl, M.}
\newblock \bibinfo{title}{Dynamical topological order parameters far from
  equilibrium}.
\newblock \emph{\bibinfo{journal}{Phys. Rev. B}} \textbf{\bibinfo{volume}{93}},
  \bibinfo{pages}{085416} (\bibinfo{year}{2016}).

\bibitem{Heyl2013}
\bibinfo{author}{Heyl, M.}, \bibinfo{author}{Polkovnikov, A.} \&
  \bibinfo{author}{Kehrein, S.}
\newblock \bibinfo{title}{Dynamical quantum phase transitions in the
  transverse-field ising model}.
\newblock \emph{\bibinfo{journal}{Phys. Rev. Lett.}}
  \textbf{\bibinfo{volume}{110}}, \bibinfo{pages}{135704}
  (\bibinfo{year}{2013}).

\bibitem{Markus2018}
\bibinfo{author}{Heyl, M.}
\newblock \bibinfo{title}{Dynamical quantum phase transitions: a review}.
\newblock \emph{\bibinfo{journal}{Reports on Progress in Physics}}
  \textbf{\bibinfo{volume}{81}}, \bibinfo{pages}{054001}
  (\bibinfo{year}{2018}).

\bibitem{Jurcevic2017}
\bibinfo{author}{Jurcevic, P.} \emph{et~al.}
\newblock \bibinfo{title}{Direct observation of dynamical quantum phase
  transitions in an interacting many-body system}.
\newblock \emph{\bibinfo{journal}{Phys. Rev. Lett.}}
  \textbf{\bibinfo{volume}{119}}, \bibinfo{pages}{080501}
  (\bibinfo{year}{2017}).

\bibitem{Zhang2017}
\bibinfo{author}{Zhang, J.} \emph{et~al.}
\newblock \bibinfo{title}{Observation of a many-body dynamical phase transition
  with a 53-qubit quantum simulator}.
\newblock \emph{\bibinfo{journal}{Nature}} \textbf{\bibinfo{volume}{551}},
  \bibinfo{pages}{601} (\bibinfo{year}{2017}).

\bibitem{Bernien2017}
\bibinfo{author}{Bernien, H.} \emph{et~al.}
\newblock \bibinfo{title}{Probing many-body dynamics on a 51-atom quantum
  simulator}.
\newblock \emph{\bibinfo{journal}{Nature}} \textbf{\bibinfo{volume}{551}},
  \bibinfo{pages}{579} (\bibinfo{year}{2017}).

\bibitem{Flaschner2018}
\bibinfo{author}{Fl\"aschner, N.} \emph{et~al.}
\newblock \bibinfo{title}{Observation of dynamical vortices after quenches in a
  system with topology}.
\newblock \emph{\bibinfo{journal}{Nat. Phys.}} \textbf{\bibinfo{volume}{14}},
  \bibinfo{pages}{265} (\bibinfo{year}{2018}).

\bibitem{Vajna2015}
\bibinfo{author}{Vajna, S.} \& \bibinfo{author}{D\'ora, B.}
\newblock \bibinfo{title}{Topological classification of dynamical phase
  transitions}.
\newblock \emph{\bibinfo{journal}{Phys. Rev. B}} \textbf{\bibinfo{volume}{91}},
  \bibinfo{pages}{155127} (\bibinfo{year}{2015}).

\bibitem{Guo2018}
\bibinfo{author}{Guo, X.-Y.} \emph{et~al.}
\newblock \bibinfo{title}{Observation of dynamical quantum phase transition by
  a superconducting qubit simulation}.
\newblock \emph{\bibinfo{journal}{arXiv:1806.09269 [quant-ph]}}
  (\bibinfo{year}{2018}).

\bibitem{Wang2018}
\bibinfo{author}{Wang, K.} \emph{et~al.}
\newblock \bibinfo{title}{Simulating dynamic quantum phase transitions in
  photonic quantum walks}.
\newblock \emph{\bibinfo{journal}{arXiv:1806.10871 [quant-ph]}}
  (\bibinfo{year}{2018}).

\bibitem{Smale2018}
\bibinfo{author}{Smale, S.} \emph{et~al.}
\newblock \bibinfo{title}{Observation of a dynamical phase transition in the
  collective heisenberg model}.
\newblock \emph{\bibinfo{journal}{arXiv:1806.11044 [quant-ph]}}
  (\bibinfo{year}{2018}).

\bibitem{Tian2018}
\bibinfo{author}{Tian, T.} \emph{et~al.}
\newblock \bibinfo{title}{Direct observation of pancharatnam geometric phase in
  a quenched topological system}.
\newblock \emph{\bibinfo{journal}{arXiv:1807.04483 [quant-ph]}}
  (\bibinfo{year}{2018}).

\end{thebibliography}


~\\
\noindent
{\bf\large Acknowledgments}

\noindent
This work was supported by National Key Research and Development Program of China (Nos. 2017YFA0304100, 2016YFA0302700), the National Natural Science Foundation of China (Nos. 61327901, 11474267, 11774335, 61322506), Key Research Program of Frontier Sciences, CAS (No.QYZDY-SSW-SLH003), the Fundamental Research Funds for the Central Universities (No.WK2470000026), the National Postdoctoral Program for Innovative Talents (No. BX201600146), China Postdoctoral Science Foundation (No. 2017M612073), and Anhui Initiative in Quantum Information Technologies (Grant No. AHY020100, AHY060300,). J.C.B. acknowledges financial support from the German Research Foundation (DFG) through the Collaborative Research Centre SFB 1143 and M. H. by the Deutsche Forschungs-gemeinschaft via the Gottfried Wilhelm Leibniz Prize
program.

\newpage

\newpage
\textbf{Caption for Fig.\,\ref{fig:s0}}: (a) shows the quenching strategy in phase diagram: starting from a ground state of Hamiltonian with flat band ($\theta_2 = \pi$) in trivial phase and ending in one non-trivial phase (pentagram with $\theta_1=8\pi/9$ and $\theta_2 = -\pi/3$). The energy band (theoretical) and the initial state (black point for theoretical expectation and arrow for experiment) are presented in (b). We fit the experimental data in rank 2 for revealing the decoherence, and the purity for each step is given in (c). In (d), we show how to extract the dynamical phase with the full knowledge of the wave-function for each step. The imaginary part of the Loschmidt amplitude is presented in the top, with three cases $k=1,~11,~21$ showing in the bottom. We read out its amplitude and period by fitting the measured results (circle points) to a trigonometric function. Density plot of the associated PGP $\phi_k^G(t)$ is given in (e), from left to right: theoretically consideration in momentum space (continuous time evolution), theoretically simulation of the QW (discrete time evolution) and our experimental results. The exact critical time is calculated from the continuous time evolution which reads $t_c = 1.513$ and predicts the first occurrence of DQPT. Experimentally measured DTOP is presented in (f) by the opaque bars (blue line is the theoretical prediction numerically calculated in momentum space (continuously) and transparent bars are predictions from the simulation of QW. Vertical dashed lines gives the critical times for each occurrence of DQPT. In the bottom, we present the rate function $\lambda (t)$ with the red line 
(obtained in continuous simulation) and the experimental measured values with points. Each non-analyticities predicts the occurrence of DQPT.

\newpage

\end{document}